# Visual Secret Sharing Scheme using Grayscale Images

Sandeep Katta
Department of Computer Science, Oklahoma State University
Stillwater, OK 74078

**ABSTRACT:** Pixel expansion and the quality of the reconstructed secret image has been a major issue of visual secret sharing (VSS) schemes. A number of probabilistic VSS schemes with minimum pixel expansion have been proposed for black and white (binary) secret images. This paper presents a probabilistic (2, 3)-VSS scheme for gray scale images. Its pixel expansion is larger in size but the quality of the image is perfect when it's reconstructed. The construction of the shadow images (transparent shares) is based on the binary OR operation.

## 1. INTRODUCTION

Secret sharing techniques belong to the larger area of information hiding that includes watermarking [1]-[8]. In secret sharing, random looking shares when brought together recreate the secret. In recursive secret sharing, the shares themselves have components defined at a lower recursive level [3]-[6]. The injection of the random bits in the shares may be done conveniently using d-sequences [9]-[11] or other random sequences.

A grayscale image is an image in which the value of each single pixel is a sample, that is, it carries only intensity information. The darkest possible shade is black, which is the total absence of transmitted or reflected light and the lightest possible shade is white.

According to their physical characteristics, different media use different ways to represent the color level of images. The computer screen uses the electric current to control lightness of the pixels. The diversity of the lightness generates different color levels. The general printer, such as dot matrix printers, laser printers, and jet printers can only control a single pixel to be printed (black pixel) or not to be printed (white pixel), instead of displaying the gray level. As such, the way to represent the gray level of images is to use the density of printed dots. The method that uses the density of the net dots to simulate the gray level is called "halftone" and transforms an image with gray level into a binary image before processing. Every pixel of the transformed halftone image has only two possible color levels (black or white). Because human eyes cannot identify too tiny printed dots and, when viewing a dot, tend to cover its nearby dots, we can simulate different gray levels through the density of printed dots, even though the transformed image actually has only two colors – black and white.

Visual secret sharing (VSS) schemes [1]-[4],[12] have been proposed only with black and white (binary) images. Several schemes for grayscale images [14] and for color images [13], [15] have been proposed. However, these earlier works result in a decrypted image of reduced quality. I here propose a new gray-level visual cryptography scheme and the image quality in this proposed scheme is better than anything and provides high quality images including that of perfect (original) quality to be reconstructed. The generation of the shadow images is based on Boolean operations, and the reconstruction operation uses OR, as in other VSS schemes.

## 2. PROPOSED APPROACH

In the proposed scheme I convert each grayscale block into a binary block. First of all each pixel value in a grayscale block is transformed into binary representation. For example take a grayscale block and transform into binary blocks.

$$\begin{bmatrix} 111 & 159 & 20 \\ 254 & 10 & 198 \\ 40 & 215 & 100 \end{bmatrix}$$

Its corresponding binary blocks are as follows:

[0 1 1 0 1 1 1 1]              [1 0 0 1 1 1 1 1]              [0 0 0 1 0 1 0 0];

[1 1 1 1 1 1 1 0]              [0 0 0 0 1 0 1 0]              [1 1 0 0 0 1 1 0];

[0 0 1 0 1 0 0 0]              [1 1 0 1 0 1 1 1]              [0 1 1 0 0 1 0 0].

Take each binary block and go for different possible combinations of that block, and try to design the block into different shares. For example take a grayscale block and divide the block into shares and apply the above scheme.

## 2.1 Two-out-of-Three Scheme using Grayscale Images

This proposed scheme is totally different from that of previous schemes. Here I design the shares such a way that when combining any two shares will reveal the original bit information, but not the whole share just half of each single share will give me high quality image when reconstructed. I will explain this scheme by taking a value from the grayscale block and divide that value into shares.

254: [1 1 1 1 1 1 1 0]

|          | 1$^{st}$ half   | 2$^{nd}$ half   |
|---------:|:---------------:|:---------------:|
| Share1:  | 0 1 0 1 0 1 0 0 | 1 1 0 1 1 0 1 0 |
| Share2:  | 1 0 1 0 1 0 1 0 | 1 1 1 0 1 1 1 0 |
| Share3:  | 0 0 1 0 0 1 0 0 | 1 0 0 1 0 1 0 0 |

**Table-1:** Grayscale bits are transformed into Binary bits

Share1 (1$^{st}$ half): 0 1 0 1 0 1 0 0    Share3 (1$^{st}$ half): 0 0 1 0 0 1 0 0
Share2 (1$^{st}$ half): 1 0 1 0 1 0 1 0    Share1 (2$^{nd}$ half): 1 1 0 1 1 0 1 0
$\overline{\phantom{xxxx}1\ 1\ 1\ 1\ 1\ 1\ 1\ 0} = 254$    $\overline{\phantom{xxxx}1\ 1\ 1\ 1\ 1\ 1\ 1\ 0} = 254$

Share2 (2$^{nd}$ half): 1 1 1 0 1 1 1 0
Share3 (2$^{nd}$ half): 1 0 0 1 0 1 0 0
$$\overline{1\ 1\ 1\ 1\ 1\ 1\ 1\ 0} = 254$$

     Combining any two half shares will give me exact bit and by doing the same procedure for the whole grayscale block gives me perfect high quality image when reconstructed without any loss of contrast.

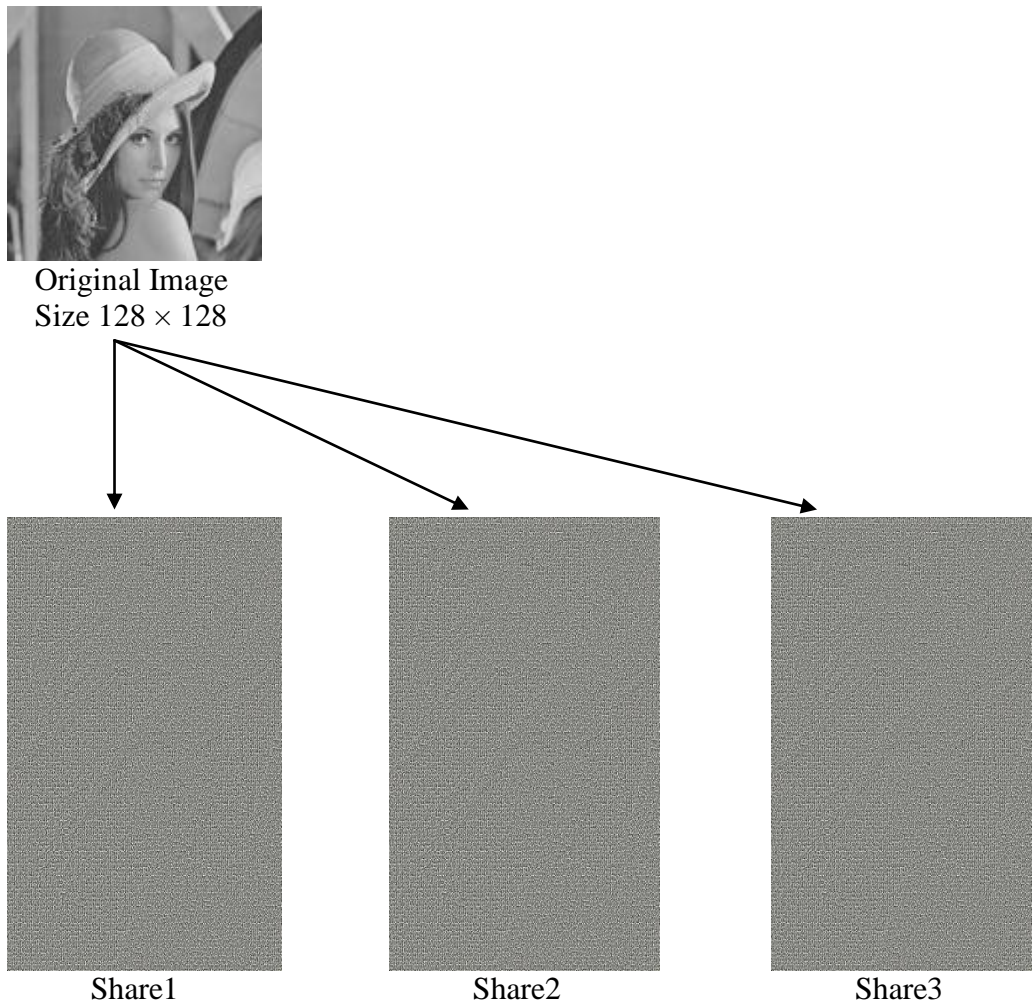

**Figure-1**: Generating three separate shared transparencies for gray-level visual cryptography

The beauty of this scheme is, when you combine the direct shares you can't see a perfect gray-scale image only when you combine the half shares, the original quality of the image will be revealed without any loss of generality.

**2.2    Simulation Results for 2-out-of-3 Visual Secret sharing Scheme using Grayscale**

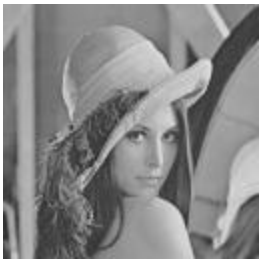    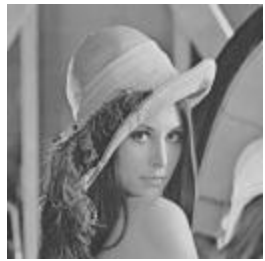    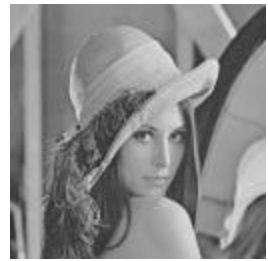

Share1(1$^{st}$ half)          Share3(1$^{st}$ half)          Share2(2$^{nd}$ half)

&                              &                              &

Share2(1$^{st}$ half)          Share1(2$^{nd}$ half)          Share3(2$^{nd}$ half)

**Figure-2**: Stacking of gray-level visual cryptography

### 3. CONCLUSION

This paper proposes a probabilistic 2-out-of-3 visual secret sharing scheme for grayscale images and gives a high quality images that of perfect (original) quality to be reconstructed. I am currently investigating to modify the grayscale secret sharing scheme in to most efficient way. In my scheme the quality of the image is maintained perfectly without any loss of generality but the size of the shadow is increased drastically, which represents the pixel expansion problem.

# REFERENCES

1. Kafri, O and Keren, E. 1987. Encryption of pictures and shapes by random grids. *Optics Letters* 12: 377-379.

2. Naor, M. and Shamir, A. 1995. Visual cryptography. *Advances in Cryptography-Eurocrypt*, 950: 1-12.

3. Gnanaguruparan, M. and Kak, S. 2002. Recursive hiding of secrets in visual cryptography. *Cryptologia* 26: 68-76.

4. Parakh, A. and Kak, S. 2008. A recursive threshold visual cryptography scheme. *Cryptology ePrint Archive, Report* 2008/535.

5. Parakh, A. and Kak, S. 2010. A tree based recursive information hiding scheme. Proceedings of IEEE ICC 2010 – *Communication and Information System Security Symposium* ('ICC'10 CISS'), May 23-27, Cape Town, South Africa.

6. Parakh, A. and Kak, S. 2011. Space efficient secret sharing for implicit data security. Information Sciences 181: 335-341.

7. Mandhani, N. and Kak, S. 2005. Watermarking using decimal sequences. Cryptologia, vol 29, pp. 50-58; arXiv: cs.CR/0602003

8. Penumarthi, K. and Kak, S. 2006. Augmented watermarking. Cryptologia, vol. 30, pp 173-180.

9. Kak, S. and Chatterjee, A. 1981. On decimal sequences. IEEE Transactions on Information Theory, IT-27: 647 – 652.

10. Kak, S. 1985. Encryption and error-correction coding using D sequences. IEEE Transactions on Computers, C-34: 803-809.

11. Kak, S. 1987. New results on d-sequences. Electronics Letters, 23: 617.

12. Sandeep, K. 2010. Recursive information hiding in visual cryptography *Cryptology ePrint Archive, Report* 2010/283.

13. Muecke, I. 1999. Greyscale and Colour Visual Cryptography, Thesis of degree of Master of Computer Science, Dalhouse University – Daltech.

14. Blude, C., De Santis, A. and Naor, M. 2000. Visual cryptography for grey level images, *Information Processing Letters*, Vol. 27, pp. 255-259.

15. Hou, Y. C. Visual cryptography for color images. *Pattern Recognition*, 36:1619-1629, 2003.